\documentclass[11pt,oneside]{amsart}
\theoremstyle{plain}
\newtheorem{thm}{Theorem}[section]
\numberwithin{equation}{section} 
\numberwithin{figure}{section} 

\newtheorem{fact}{Proposition}[section]
\newtheorem{lemma}{Lemma}[section]

{\noindent{\hfill\mbox{\rule{.5em}{.5em}}\,}\par\medskip}

\def\PRD{Phys.\ Rev.\ D}

\def\PLB{Phys.\ Lett.\ B}

\newcommand{\bref}[1]{{\bf \ref{#1}}}

\newcommand{\qcommut}[2]{[#1,#2]_\star}

\newcommand{\pb}[2]{\left\{{}#1{},{}#2{}\right\}}

\newcommand{\gh}[1]{{\rm gh}(#1)}
\newcommand{\p}[1]{{\rm p}(#1)}
\renewcommand{\deg}[1]{{\rm deg}(#1)}
\renewcommand{\dim}[1]{{\rm dim}(#1)}

\def\bar{\overline}
\newcommand{\func}[1]{{{\mathcal C}^\infty}{(#1)}}             %
\def\tensor{\otimes}
\def\d{\partial}

\newcommand{\dl}[1]{\displaystyle\frac{{\d}}{\d #1}}
\newcommand{\ddl}[2]{\displaystyle\frac{{\d #1}}{\d #2}}
\def\half{\frac{1}{2}}

\def\cP{{\mathcal P}}
\def\cc{{\mathcal C}}

\def\cT{{\rm T}}
\newcommand{\formal}[1]{{{\mathcal F}}{(#1)}}             %
\def\mod{{\mathcal T}^*_\rho}
\def\modbar{{\mathcal T}^*_{\bar\rho}}

\def\manL{{\mathcal L}}
\def\manM{{\mathcal M}}

\def\aA{{ \mathfrak A}}
\def\E{{ \mathcal E}}
\def\P{{ \bf P}}
\def\G{{ \bf G}}

\begin{document}
\thispagestyle{empty}
\vfuzz1.6pt
\addtolength{\baselineskip}{4pt}
\addtolength{\parskip}{2pt}
\raggedbottom

{\hfill{\lowercase{\tt hep-th/0003114}}\\[12pt]}
\title{Fedosov Deformation Quantization as a BRST Theory.}

\begin{center}
\LARGE{Fedosov Deformation Quantization as a BRST Theory.}
\end{center}

\bigskip

\begin{center}
{\large{M.A.~Grigoriev$^1$ and S.L.~Lyakhovich$^2$}}
\end{center}

\begin{center}
\textit{$^1$Lebedev Physics Institute, Russian Academy of Sciences}\\
\textit{$^2$Department of Physics, Tomsk State University}
\end{center}

\vspace{10mm}

\begin{center}
\begin{minipage}{150mm}
\footnotesize{ The relationship is established between the Fedosov
deformation quantization of a general symplectic manifold and the BFV-BRST
quantization of constrained dynamical systems.  The original symplectic
manifold $\manM$ is presented as a second class constrained surface in the
fibre bundle $\mod\manM$ which is a certain modification of a usual
cotangent bundle equipped with a natural symplectic
structure.  The second class system is converted into the first class one
by continuation of the constraints into the extended manifold, being a
direct sum of $\mod\manM$ and the tangent bundle $T\manM$.  This extended
manifold is equipped with a nontrivial Poisson bracket which naturally
involves two basic ingredients of Fedosov geometry:  the symplectic
structure and the symplectic connection.  The constructed first class
constrained theory, being equivalent to the original symplectic manifold,
is quantized through the BFV-BRST procedure.  The existence theorem is
proven for the quantum BRST charge and the quantum BRST invariant
observables.  The adjoint action of the quantum BRST charge is
identified with the Abelian Fedosov connection while any observable,
being proven to be a unique BRST invariant continuation for the values
defined in the original symplectic manifold, is identified with the
Fedosov flat section of the Weyl bundle.  The Fedosov fibrewise star
multiplication is thus recognized as a conventional
product of the quantum BRST invariant observables.}
\end{minipage}
\end{center}

\bigskip

\bigskip

\section{Introduction}
Different trends are recognized among the approaches to quantization
of systems whose classical mechanics is based on the Poisson
bracket. In physics, the quantization strategy evolves, in a sense, in the
opposite direction to the main stream developing
in mathematics. {}From the physical viewpoint, the phase manifold is
usually treated as a constraint surface in a flat manifold
or in a manifold whose geometric structure
is rather simpler than that of constraint surface.  And the efforts are
not directed to reduce the dynamics on the curved shell before
quantization.  The matter is that the physical models should usually
possess an explicit relativistic covariance and space-time locality,
whereas the reduction to the constraint surface usually breaks both.  So,
the main trend in physics is to quantize the system as it originally
occurs, i.e. with constraints.  The reduction is achieved in quantum
theory by means of restrictions imposed to the class of admissible
observables and states.  The most sophisticated quantization scheme
developed in this direction is the BFV method \cite{[BFV]} (for review see
\cite{[HT]}) based on the idea of the BRST symmetry. The method allows, in
principle, to quantize any first class constraint theory, with the
exception of the special case of the so-called infinitely reducible
constraints. As to the second class constrained theories, various methods
are known  to  adopt the BFV-BRST approach for the case.  In this paper we
turn to the idea to convert the second class theory into the first class
extending the phase manifold by extra degrees of freedom which are going
to be eventually gauged out by the introduced gauge symmetry related to
the effective first class constraints.  A number of the general conversion
schemes is known today \cite{[FSh],[BF],[BF87],[BT]}.  The conversion ideas
are widely applied in practical physical problems concerning quantization of
the second class constrained systems.

\noindent
The mathematical insight into the quantization problem always starts
with the reduced Poisson manifold where the constraints, if they could
originally present, have already been resolved.
The general concept of the deformation quantization
was introduced in Ref. \cite{[Berezin],[BFFLS]}.
The existence of the star product on the
general symplectic manifold was proven in Ref. \cite{[DWL]}
where the default was ascertained for the cohomological obstructions
to the deformation of the associative multiplication.

\noindent
Independently, Fedosov suggested the explicit construction of the
star product on any symplectic manifold \cite{[Fedosov-JDG]}
(see also the subsequent book
\cite{[Fedosov-book]}).  Now the general
statement regarding the existence of the star product for the most general
Poisson manifold is established by
Kontsevich~\cite{[Kontsevich]}.  Recently the Kontsevich
quantization formula was also supplied with an interesting physical
explanation~\cite{[Felder]}.

\noindent
However, in the case of symplectic manifolds
the Fedosov construction of the star product
seems to be most useful in applications.  The advantage
is in the explicit description of the algebra of quantum
observables.  In the Fedosov approach the quantum
observable algebra is the space of the flat sections of the Weyl
algebra bundle over the symplectic manifold, with the
multiplication being the fibrewise Weyl product.  The Fedosov
star-product allows a generalisation to the case of super-Poison
bracket~\cite{[Bord-super]}.

\noindent
Mention that the deformation quantization structures are
coming now to the gauge field theory not only as a tool
of quantizing but rather as the means of constructing
new classical models, e.g. gauge theories on
noncommutative spaces~\cite{[Connes]} and higher-spin
interactions~\cite{[Vasiliev]}.  The recent developments
have also revealed a deep relationship between the strings
and the Yang-Mills theory on the noncommutative
spaces~\cite{[SW],[CDS]}.

\noindent
The BRST approach to the quantization of the systems with the
geometrically nontrivial phase space was initiated by Batalin and
Fradkin who suggested to present the original symplectic manifold as a
second class constraint surface embedded into the linear symplectic
space \cite{[BF89]}\footnote{The global geometric properties
of this embedding were not in the focus of the original papers
\cite{[BF89],[FL],[BFF]}.  In the present paper we suggest a slightly
different constrained embedding of the symplectic manifold with an explicit
account for the global geometry, although our basic goal is beyond
the geometry of the embedding itself.}

\noindent
{}From the viewpoint of the BFV method, the question of deformation
quantization of general symplectic manifold was considered
in Ref.~\cite{[FL]} where one could actually observe (although it was not
explicitly mentioned about in the paper) that the generating equations
for the Abelian conversion~\cite{[BT]}, being applied to the
embedding of the second class constraints of the
Ref.\cite{[BF89],[BFF]}, naturally involve the characteristic
structures of the Fedosov geometry:  the symmetric symplectic
connection and the curvature.

\noindent
However, according to our knowledge,  the relationship has not been
established yet between the second class constraint approach of
\cite{[BF89],[FL],[BFF]} and the Fedosov construction.

\noindent
In this paper we show that the Fedosov quantization scheme
can be completely derived from the BFV-BRST quantization of the
constrained dynamical systems.

\noindent
First the symplectic manifold $\manM$ is extended to the fibre bundle
$\mod\manM$, being a certain modification of the usual cotangent
bundle, which still carries the canonical symplectic structure.  The
original manifold $\manM$ is identified to the second class
constrained surface in $\mod\manM$.  This allows to view the
Poisson bracket on the base manifold as the Dirac bracket associated
to the second class constraints.  Further, the second class
constraints are converted into the first class ones in spirit of the
Abelian conversion procedure~\cite{[BT]}.  In the case at hand, we
choose the conversion variables to be the coordinates on the fibres
of the tangent bundle over the symplectic manifold.  The phase space
of the converted system, in distinction to the direct application of
the conventional conversion scheme~\cite{[BT]} exploited in the
Ref.~\cite{[FL]}, is equipped with a natural nonlinear symplectic
structure.  This symplectic structure involves the initial symplectic
form and a symmetric symplectic connection.  Remarkably,
these structures are known as those determining the so called
Fedosov geometry~\cite{[GRS]}.  In its turn the Jacobi identity
for the Poisson bracket, being defined in this extended manifold, encodes
all the respective compatibility conditions for the Fedosov manifold. So,
the embedding and converting procedure make the relationship
transparent between the constrained Hamiltonian dynamics and
Fedosov's geometry.

\noindent
We quantize the resulting  gauge invariant system, being globally
equivalent to the original symplectic one, according to the standard BFV
quantization prescription.  As the extended phase space of the BFV
quantization is a geometrically nontrivial symplectic manifold, it is a
problem to quantize it directly.  Fortunately, to proceed with the
BFV scheme when the constraints have the special structure as in the case
in hand, one needs to define only the quantization of
some subalgebra of functions.  Namely, we consider subalgebra $\aA$
of functions at most linear in the momenta which is closed w.r.t.
the associative multiplication and the Poisson bracket.  This
subalgebra contains all the BRST observables, BRST charge $\Omega$ and
the ghost charge.  Unlike the entire algebra of functions on the
extended phase space, the construction of the star-multiplication
in $\aA$ is evident in this case.

\noindent
At the quantum level we arrive at the quantum BRST charge
$\Omega$ satisfying the quantum nilpotency condition
$\qcommut{\Omega}{\Omega}=0$.  The algebra of quantum
observables is thus the
zero-ghost-number cohomology of $Ad~\Omega$.  This algebra, being viewed
as a vector space, is isomorphic to the algebra
of classical observables. The noncommutative product
from the algebra of quantum BRST observables is carried over
to the space of functions on the symplectic manifold
giving a deformation quantization.

\noindent
This approach allows us to identify all the basic structures of
Fedosov's method as those of the BRST theory. In particular,
the auxiliary variables $y^i$ (which appear in Ref \cite{[Fedosov-JDG]}
as the generators of Weyl algebra) turns out to be the conversion
variables, the basic
one-forms $dx^i$ on the symplectic manifold should be identified with the
ghost variables associated to the converted constraints.  Further, the
Fedosov flat connection $D$ is the adjoint action of the quantum BRST charge
$\Omega$; the flat sections of the Weyl bundle is thus nothing but the BRST
cohomology.
Under this identification, the Fedosov quantization statements
regarding the existence of the Abelian connection, lift of
the functions from the symplectic manifold to the flat sections
of the Weyl bundle can be recognized as the standard existence
theorems of the BRST theory.

\section{ Representation of a general symplectic
manifold as a constrained Hamiltonian dynamics.}

In this section we first represent a general symplectic
manifold $\manM$, \, $\dim{\manM}=N$ as a second class
constraint surface embedded into the fibre bundle $\mod\manM$, \,
$\dim{\mod\manM}=2N$ being equipped with globally defined canonical
symplectic structure. Next we develop the procedure to convert the
second class constraints into the first class ones extending the
manifold $\mod\manM$ to the direct sum  $ \mod\manM \oplus T \manM$
which possesses a nontrivial Poisson structure.  This structure
generates, in a sense, all the structure relations of the
symplectic geometry of the original symplectic manifold $\manM$.  The
extra degrees of freedom introduced with this embedding are
effectively gauged out due to the gauge symmetry related to the
effective first class constraints.  And finally we construct the
classical BRST embedding for the effective first class system, which
serves as a starting point for the BFV-BRST quantization of the
symplectic manifold, that is done in the next section.

\subsection{Second class constraint formulation of the
symplectic structure}\label{subsec:FM}
Let $\manM$ be the symplectic manifold with symplectic form $\omega$.
Denote by $\pb{\cdot \,}{\cdot \,}_\manM $ the respective
Poisson bracket on $\manM$.  Let $x^i$ be a local coordinate
system on $\manM$.  In the local coordinates
the symplectic form and the Poisson bracket read as
\begin{equation}
\omega=\omega_{ij}(x)d x^i\wedge d x^j \,, \quad d \omega =0 ,
\label{formM}
\end{equation}
\begin{equation}
\pb{a(x)\,}{b(x)\,}_\manM=
\omega^{ij}(x)\frac{\d a(x)}{\d x^i}\frac{\d b(x) }{\d x^j}\,,\quad
\omega^{ij}\omega_{jk}=\delta^i_k.
\label{PBM}
\end{equation}

\noindent
Let $\Gamma$ be a symmetric symplectic connection on $\manM$,
which always exists (for details of the geometry based on this
connection see \cite{[GRS]}).
In the local coordinates $x^i$ we have
\begin{equation}
\label{eq:coefficients}
\dl{x^i}\omega_{jk}-\Gamma^l_{ij}\omega_{lk}-\Gamma^l_{ik}\omega_{jl}=0\,,
\qquad \nabla_i(\dl{x^j})=\Gamma^k_{ij}\dl{x^k}\,.
\end{equation}
Introduce a curvature tensor $R^k_{l;ij}$ of $\Gamma$ by
$R^k_{l;ij}\dl{x^k}=[\nabla_i,\nabla_j]\dl{x^l}$.
In the local coordinates it reads
\begin{equation}
  \label{eq:riemann}
R^k_{l;ij}=\d_i\Gamma^k_{jl}+\Gamma^n_{jl}\Gamma^k_{in}-
\d_j\Gamma^k_{il}-\Gamma^n_{il}\Gamma^k_{jn}\,.
\end{equation}

\noindent
In the symplectic geometry it is convenient to use the coefficients
$\Gamma_{ijk}$ defined by $\Gamma_{ijk}=\omega_{in}\Gamma^n_{jk}$
and $R_{kl;ij}=\omega_{kn}R^k_{l;ij}$.  The curvature
tensor $R_{kl;ij}$ satisfies corresponding Bianchi
identities:
\begin{equation}
\label{Bianchi}
\nabla_m R_{kl;ij} + \ldots = 0
\end{equation}

\noindent
The following properties are known of the symmetric symplectic
connection (see e.g.~\cite{[GRS]}): $\Gamma_{ijk}$ is total symmetric
in each Darboux coordinate system and the curvature tensor has the symmetry
property
\begin{equation}
        R_{kl;ij}=R_{lk;ij}\,.
\end{equation}
This fact could be immediately seen by choosing a coordinate system
where $\omega_{ij}$ are constant.

\noindent
Consider an open covering of $\manM$.  In each
domain $U_\alpha$ the symplectic form can be represented as
\begin{equation} \label{eq:rho} \omega=d\rho^\alpha\,, \qquad
        \omega_{ij}=\d_i\rho^\alpha_j-\d_j\rho^\alpha_i\,,
\end{equation}
where $\rho^\alpha=\rho^\alpha_i d x^i$ is the symplectic potential
in $U_\alpha$.  In the overlapping $U_{\alpha \beta}=U_\alpha \cap
U_\beta$ we have
\begin{equation}
  \rho^\alpha-\rho^\beta=\phi^{\alpha\,\beta}\,, \qquad
  d\phi^{\alpha\beta}=0\,.
\end{equation}
The transition 1-forms $\phi^{\alpha\beta}$ obviously satisfies
\begin{equation}
\phi^{\alpha\beta}+\phi^{\beta\alpha}=0\,,\qquad
\phi^{\alpha\beta}+\phi^{\beta\gamma}+\phi^{\gamma\alpha}=0\,.
\end{equation}
in the overlappings $U_\alpha\cap U_\beta$ and
$U_\alpha\cap U_\beta \cap U_\gamma$ respectively.

\noindent
Given an atlas $U_\alpha$ and the symplectic
potential $\rho^\alpha$ defined in each domain $U_\alpha$
one can construct an affine bundle $\mod\manM$
over $\manM$. Namely, for each domain $U_\alpha$ with
the local coordinates $x_\alpha^i$ (index $\alpha$ indicates
that $x^i_\alpha$ are the coordinates on $U_\alpha$) choose
the fibre to be $R^N$ ($N=\dim{\manM}$) with the coordinates
$p^\alpha_i$.  In the overlapping $U_\alpha \cap U_\beta$ we
prescribe the following transition law
\begin{equation}
\label{eq:transition}
  p^\alpha_i=p^\beta_j \frac{\d x^j_\beta}{\d x^i_\alpha}
  +\phi^{\alpha \beta}_i\,.
\end{equation}
(summation over the repeating indices $\alpha,\beta,\ldots$
is not implied).  Here the coefficients $\phi^{\alpha \beta}_i$ of
the 1-form $\phi^{\alpha \beta}$ are introduced by
$\phi^{\alpha \beta}=\phi^{\alpha \beta}_i dx^i_\alpha$.
It is easy to check that the transition law~\eqref{eq:transition}
satisfies standard conditions in the overlapping of two and three
domains and thus it determines $\mod\manM$ as a bundle.

\noindent
The difference between usual cotangent bundle $T^*\manM$ and
$\mod\manM$ is that the structure group of the former is $GL(N,R)$
while that of the later is a group of affine transformations of $R^N$.

\noindent
As the transformation law \eqref{eq:transition} of the
variables $p_i$ differs from that of the coordinates on
the fibres of the standard cotangent bundle by a closed
1-form only, then $\mod\manM$ is also equipped with the
canonical symplectic form $d p_i \wedge d x^i$. In particular,
the corresponding Poisson bracket has also the canonical form
\begin{equation}
  \label{eq:FMPB}
  \pb{f}{g}=\frac{\d f}{\d x^i}\frac{\d g}{\d p_i}-
  \frac{\d f}{\d p_i}\frac{\d g}{\d x^i}\,.
\end{equation}

\noindent
An important feature of this construction is that the
surface $\manL$ defined by the equations
\begin{equation}
\label{eq:theta}
\theta_i (x,p)\, \equiv \, \rho_i \, - \, p_i \,=0 \,,
\end{equation}
is a submanifold in $\mod\manM$. Indeed, these equations can be
considered as those determining the smooth section of
$\mod\manM$.  Moreover, considered as a manifold,
$\manL$ is isomorphic to the original manifold $\manM$.  Indeed, $\manL$
is a section of the bundle $\mod\manM$ and $\manM$ is a base of
$\mod\manM$; the projection $\pi\,:\,\mod\manM \to
\manM$ to $\manL \subset \mod\manM$ obviously establishes an isomorphism
between $\manM$ and $\manL$.  Note also that quantities $\theta_i$ transforms
as coefficients of a 1-form.

\noindent
{}From the viewpoint of the Hamiltonian constrained
dynamics, $\theta_i$ are the second class constraints, as their Poisson
brackets in $\mod\manM$ form an invertible matrix
\begin{equation}
\pb{\theta_i}{\theta_j}=\omega_{ij}
\label{thetatheta}
\end{equation}
The Dirac bracket in $\mod\manM$, being
built of the constraints \eqref{eq:theta},
\begin{equation}
\pb{f(x,p)}{g(x,p)}_D \equiv \pb{f(x,p)}{g(x,p)} -
\pb{f(x,p)}{\theta_i}\omega^{ij}\pb{\theta_j}{g(x,p)}
\label{eq:DB}
\end{equation}
can be considered as Poisson bracket defined
on the constraint surface $\manL$.  As the Dirac bracket
is nondegenerate on the constraint surface $\manL$, the
latter is a symplectic manifold.  One can see that
$\manL$ is isomorphic to $\manM$ when each one is
considered as a symplectic manifold.
Indeed, any function $f(x,p)$ on $\mod\manM$ can be reduced on $\manL$ to
the function $f_{0}(x)=f(x,p)|_{p_i=\rho_i(x)}$, while the function $f_{0}(x)$
can be understood as defined on the original manifold $\manM$.
The Dirac bracket \eqref{eq:DB} between any functions $f(x,p), \, g(x,p)$
coincides on the constraint surface $\manL$ determined by constraints
\eqref{eq:theta} to the Poisson bracket
between their projections to $\manM$:

\begin{equation}
\begin{split}
{{\{ f(x,p) , g(x,p) \}_D }{|}}_{\displaystyle p_i = \rho_i(x)}
=& \{ f_0(x) , g_0 (x) \}_{\manM} \, ,\\
f_0 (x) = f(x,p)|_{\displaystyle p_i = \rho_i(x)}\,, \qquad  \qquad &
g_0 (x) = g(x,p)|_{\displaystyle p_i = \rho_i(x)}\,.
\end{split}
\end{equation}
This obvious fact provides the equivalence of the constrained dynamics in
$\mod\manM$ and the Hamiltonian one in $\manM$.  The quantization
problem for the symplectic manifold $\manM$ is thereby equivalent to the
quantization of the second class constrained theory in $\mod\manM$.

\subsection{Conversion to the first class}

\noindent
In this section we suggest a procedure to convert second class
constraints \eqref{eq:theta} into the first class ones. The procedure
explicitly accounts the geometry of the original manifold $\manM$,
and makes transparent the relationship between the BRST and
Fedosov's constructions.

\noindent
Now we further enlarge the phase space. Namely we embed
$\mod\manM$ into the $\mod\manM \oplus T\manM$. Let
$y^i$ be the natural coordinates on the fibres of tangent
bundle $T\manM$. In order to equip the extended phase space
$\mod\manM \oplus T\manM$ with the Poisson bracket one has
to engage an additional structure, a symplectic connection.

\noindent
In view of the properties~\eqref{eq:coefficients},
\eqref{eq:riemann} and \eqref{Bianchi} one can equip
$\mod\manM\oplus T\manM$ with a symplectic structure.  Indeed,
let the bracket operation
$\{\cdot\,,\cdot\,\}$ on $\mod\manM \oplus T\manM$
be given by
\begin{equation}
\label{eq:PB}
\begin{array}{rclrcl}
\displaystyle
\pb{x^i}{p_j}&=&\delta^i_j\,,&  \quad \quad
\pb{y^i}{y^j}&=&\omega^{ij}\,,\\ [3pt]
\pb{p_i}{y^j}&=&\Gamma^j_{il} y^l \,,& \quad \quad
\pb{p_i}{p_j}&=&\half R_{mn;\,ij}y^m y^n\,, \\ [3pt]
\pb{x^i}{x^j}&=&0\,,& \quad \quad
\pb{x^i}{y^j}&=&0\,.\\
\end{array}
\end{equation}
Then \eqref{eq:PB} is a Poisson bracket provided
$\Gamma$ is a symmetric symplectic connection.

\noindent
Considering Jacobi identities for the Poisson brackets \eqref{eq:PB}
between $y^i, y^j, p_k$; \, $y^i, p_j, p_k$ and $p_i, p_j, p_k$
one arrives at~\eqref{eq:coefficients},\eqref{eq:riemann}
and \eqref{Bianchi} respectively. In this sense, the Poisson bracket
\eqref{eq:PB} might be viewed as a generating structure for
the Fedosov geometry which is based just on the relations
\eqref{eq:coefficients}, \eqref{eq:riemann} and \eqref{Bianchi}.

\noindent
In what follows instead of smooth functions
$\func{\mod\manM \oplus T\manM}$ on $\mod\manM \oplus T\manM$, we will consider
formal power series in $y^i$ with coefficients in $\func{\mod\manM}$.  Moreover,
we restrict the coefficients to be polynomials in $p_i$.  The reason
for this is that $y^i$ serve as ``conversion variables'' and
one has to allow formal power series in $y$.   As $p_i$ play a role of
momenta, it is a usual technical restriction in physics to allow
only polynomials in $p_i$.  Thus speaking about "functions"
on $\mod\manM \oplus T\manM$ we mean sections of the appropriate
vector bundle over $\manM$.  The Poisson bracket~\eqref{eq:PB}
is well defined in the algebra of these ``functions''.

\noindent
There is a simple formula which clarifies
the geometrical meaning of this Poisson bracket:
\begin{equation}
\pb{p_i}{f(x,y)} = - \nabla_i f(x,y)\,,
\label{eq:nabla}
\end{equation}
where the function $f(x,y)$ (formal power series in $y$)
is understood in the r.h.s.  of \eqref{eq:nabla} as an inhomogeneous
symmetric tensor field on $\manM$, i.e.
\begin{equation} f(x,y)=\sum^\infty_{k=0}
f_{i_1 \ldots i_k} (x) \, y^{i_1} \ldots y^{i_k} \, , \quad \pb{f(x,y)}{p_j}
= \sum^\infty_{k=0} \bigl( \nabla_j f_{i_1 \ldots i_k} (x) \bigr) \, y^{i_1}
\ldots y^{i_k}\,.
\end{equation}

\noindent
The goal of the conversion procedure is to continue
the second class constraints $\theta_i(x,p)$ \eqref{eq:theta},
being the functions on
$\mod\manM$, into   $\mod\manM \oplus T \manM$  \,
$\theta_i \to { \cT}_i (x,p,y), \, { \cT}_i|_{y=0}=\theta_i $
in such a way that ${\cT}_i$ have to be the first class
in the extended manifold.
Thus we look for the functions $\cT_i$ such that
\begin{equation}
\label{eq:main-convers}
\pb{\cT_i}{\cT_j}=0 \,, \quad
{\cT}_i|_{y=0}= {\theta}_i\,.
\end{equation}
We also prescribe $\cT_i$ to transform as the
coefficients of the 1-form
under the change of coordinates on $\manM$,
as the original constraints $\theta_i$~\eqref{eq:theta}
have the same transformation property.  Existence of the
Abelian conversion is established by the
following\footnote{The local proof for the respective
existence theorem is known \cite{[BT]}.
However we give here a proof with a due regard
to the global geometry which is based on the Poisson
bracket~\eqref{eq:PB}.}
\begin{fact}\label{fact:existence}
Equation~\eqref{eq:main-convers} has a solution.
\end{fact}
\begin{proof}
Let us look for the solution to the equation~\eqref{eq:main-convers}
in the form of the explicit power series expansion
in the variables $y^i$.
\begin{equation}
\label{eq:expansion}
\cT_i=\sum^\infty_{r=0} \tau^r_i
\end{equation}
It turns out that it is sufficient to consider functions
$\tau^r_i,\, r \geq 1$ which do not depend on the momenta $p_i$:
\begin{equation}
\tau^0_i(x)=\theta_i\,, \quad
\tau^r_i=\tau^r_{i\, j_1\ldots
j_r}(x)y^{j_1}\ldots y^{j_r}\,.
\end{equation}
Functions $\tau^r_{i\, j_1\ldots j_r}(x)$ can
be considered as the coefficients of the tensor field
on $\manM$ that is symmetric w.r.t. all indices except
the first one.  In the zeroth and first order we
respectively have
\begin{equation}
\label{eq:zeroth-first}
\omega_{ij}+\tau^1_{i\,l} \omega^{l\,k}\tau^1_{jk}=0\,,
\qquad
\nabla_{[i}\tau^1_{j]k}+2\tau^2_{[ilk}\omega^{lm}\tau^1_{j]m}=0\, ,
\end{equation}
with $[i,j]$ standing for antisymmetrisation in $i,j$.
There is a particular solution to these equations:
\begin{equation}
\tau^1_{i\,k}=-\omega_{ik}\,,\qquad \tau^2_{ijk}=0\,.
\label{eq:tau-particular}
\end{equation}
Taking $\tau^1_{ij}=-\omega_{ij}$ one can in fact
consider more general solutions for $\tau^2_{ijk}$. In this case,
the second equation of~\eqref{eq:zeroth-first} implies that
$\Gamma_{ijk}+\tau^2_{ijk}$ are the coefficients of a symmetric
symplectic connection on $\manM$. This arbitrariness
in the solution of~\eqref{eq:zeroth-first} can be absorbed
by the redefinition of the symmetric symplectic connection
entering the Poisson bracket \eqref{eq:PB}.

\noindent
The ambiguity in $\tau^1_{il}$ might be able to reflect
additional geometrical structures on $\manM$.  As we
will see below, standard Fedosov's construction of the
star-product on $\manM$ corresponds to the ``minimal''
solution~\eqref{eq:tau-particular}. However we consider here a
general solution to \eqref{eq:zeroth-first}.

\noindent
Taking any fixed solution to Eq.~\eqref{eq:zeroth-first} for
$\tau^1_i,\tau^2_i$ one sees that in the r-th $(r\geq2)$ order in $y$
Eq.~\eqref{eq:main-convers} implies
\begin{equation} \label{eq:determin}
\pb{\tau^1_{[ i}}{\tau^{r+1}_{j ]}}+B^r_{ij}=0\,,
\end{equation}
where the quantities $B^{r}_{ij}$ are given by
\begin{equation} \begin{split}
\label{eq:B^n}
B^2_{ij}&=\nabla_{[ i}\tau^2_{ j ]}+ \pb{\tau^2_{i}}{\tau^2_{j}}+
\half R_{mn;ij}y^m y^n\,,\\
B^r_{ij}&=\nabla_{[ i} \tau^r_{j ]}+
\sum_{t=0}^{r-2} \pb{\tau_i^{r-t}}{\tau_j^{t+2}}\,, \qquad r \geq 3 \,.
\end{split}
\end{equation}
Now relations~\eqref{eq:determin} are to be considered as the equations
determining $\tau^{r+1}_i$.
We need the following
\begin{lemma}
Let the quantity $A_{ij}(x,y)$ be such that $A_{ij}+A_{ji}=0$ and
$\pb{\tau^1_i}{A_{jk}}+{\rm cycle}(i,j,k)=0$ then
there exist $C_{i}$ such that
\begin{equation}
A_{ij}=\pb{\tau^1_{[i}}{C_{j]}}\,.
\end{equation}
\end{lemma}
\noindent
The statement is an obvious generalisation of the standard
Poincare Lemma. In the case where $\tau^1_{i\,k}=-\omega_{ik}$,
it is precisely the Poincare Lemma.

\noindent
It follows from the lemma that equation
\eqref{eq:determin} has a solution iff $B_{ij}$
satisfies
\begin{equation}
\pb{\tau^1_i}{B_{jk}}+{\rm cycle}(i,j,k)=0\,.
\end{equation}
To show that it takes place let us introduce the partial sum
\begin{equation}
  \cT^s_i=\sum_{t=0}^s \tau^t_i
\end{equation}
and consider expression
\begin{equation}
  \pb{\cT^s_i}{\cT^s_j}=\sum_{t=1}^{(s-1)}
(\pb{\tau^1_{[ i}}{\tau^{t+1}_{j ]}}+
B^t_{ij})+B^s_{ij}+ \ldots \,.
\end{equation}
where $\ldots$ denote terms of order higher
than $s$ in $y^i$.  Assume that Eqs.~\eqref{eq:determin}
hold for $1 \leq r \leq s-1$.  Excluding the contribution of
order $s-1$ from the Jacobi
identity
\begin{equation}
  \pb{\cT^s_i}{\pb{\cT^s_j}{\cT^s_k}}+cycle(i,j,k)=0\,,
\end{equation}
one arrives at
\begin{equation}
  \label{eq:comp}
\pb{\tau^1_i}{B^s_{jk}}+cycle(i,j,k)=0\,.
\end{equation}
It follows from the lemma that for $r=s$ equation~\eqref{eq:determin}
considered as that on $\tau^{r+1}$  admits a
solution.  The induction implies that Eq.~\eqref{eq:main-convers}
also admits solution, at least locally.

\noindent
To show that solution exists globally we construct the
particular solution to the equation \eqref{eq:determin} for
$r=s$:
\begin{equation}
\label{eq:particular}
  \tau^{s+1}_i=
-\frac{2}{s+2}
B^s_{ij} (K^{-1})^j_m  y^m\,, \qquad K_i^j=\tau^1_{il}\omega^{lj}\,.
\end{equation}
This solution satisfies the condition
\begin{equation}
\label{eq:cond}
  \tau^{s+1}_i (K^{-1})^i_j y^j=0\,,
\end{equation}
which does not depend on the choice of
the local coordinates on $\manM$.  Given a fixed first
term $\tau^1_i$, Eq. \eqref{eq:cond} can be considered
as the condition on the solution to the
equation~\eqref{eq:main-convers}.  It is easy to see that solution to
\eqref{eq:main-convers} is unique provided the condition~\eqref{eq:cond}
is imposed.  Indeed, general solution to the Eq.~\eqref{eq:determin} is
given by
\begin{equation}
\label{eq:classical-general}
{\bar\tau}_i^{s+1}=\tau_i^{s+1}+\pb{\tau^1_i}{C^s}\,.
\end{equation}
where $\tau_i^{s+1}$ is the particular solution~\eqref{eq:particular}
and $C^s=C^s(x,y)$ is an arbitrary function. One can check
that condition~\eqref{eq:cond} implies that second term
in~\eqref{eq:classical-general} vanishes.  Choosing $\tau_i$
to satisfy~\eqref{eq:cond} in each domain $U_\alpha$ one gets
the global solution to Eq.~\eqref{eq:main-convers}.
\end{proof}

\noindent
Thus we have arrived at the first class constrained system,
with the constraints being~\eqref{eq:expansion}.  An
observable of the first class constrained system
is a function $A(x,y,p)$ satisfying
\begin{equation}
\label{eq:GI0}
\pb{\cT_i}{A}=V_i^j\cT_j\,.
\end{equation}
for some functions $V^i_j(x,y,p)$.  Observables $A(x,y,p)$ and
$B(x,y,p)$ are said equivalent iff
their difference is proportional to the constraints, i.e.
\begin{equation}
A-B=V^i\cT_i
\end{equation}
for some functions $V^i(x,y,p)$.  In each equivalence class of the
observables there is a unique representative which does not depend
on the momenta $p_i$.  Indeed, let $A(x,y,p)$ be an
observable.  Since it is a polynomial in $p$ it can be rewritten as
\begin{equation}
A=a(x,y)+a^i_1(x,y)\cT_i+\ldots\,,
\end{equation}
where $\ldots$ stands for higher (but finite) orders in
$\cT$.  It follows from~\eqref{eq:GI0} that $a(x,y)$ satisfies
\begin{equation}
\label{eq:GI} \pb{\cT_i}{a(x,y)}=0\,.
\end{equation}
Now we are going to show that the
Poisson algebra of the inequivalent observables is isomorphic
to the algebra of functions on $\manM$.
\begin{fact}
\label{fact:extention}
Eq.~\eqref{eq:GI} has a unique solution $a(x,y)$ satisfying
$a(x,0)=a_0(x)$ for any given function $a_0 \in \func\manM$
\end{fact}
\noindent
A proof is a direct analogue of that of
Proposition~\eqref{fact:existence}.

\noindent
Given two solutions $a(x,y)$ and
$b(x,y)$ of Eq.~\eqref{eq:GI} corresponding to the
boundary conditions $a(x,0)=a_0(x)$ and $b(x,0)=b_0(x)$ one
can check that
\begin{equation}
         \pb{a}{b}|_{y=0}=\pb{a_0}{b_0}_{\manM}\,.
\end{equation}
Thus the isomorphism is obviously seen between Poisson algebra
of observables of the first class theory and the Poisson algebra
of functions of symplectic manifold $\manM$.  This
shows that the constructed first class constrained
system is equivalent to the original unconstrained
system on $\manM$.

\subsection{An extended Poisson bracket and the BRST charge}
According to the BFV quantization prescription we
have to extend the phase space introducing
Grassmann odd ghost variable $\cc^i$ to each constraint
$\cT_i$ and the ghost momenta $\cP_i$ canonically conjugated
to $\cc^i$.

\noindent
We wish the ghost variables $\cc^i$ and $\cP_i$ to transform under the
change of local coordinate system on the base $\manM$ as the components
of the vector field and 1-form respectively. Thus in the intersection
$U_\alpha \cap U_\beta$ one has
\begin{equation}
\cc^i_{\alpha}=\cc^j_\beta \ddl{x^i_\alpha}{x^j_\beta}\,, \qquad
\cP^\alpha_i=\cP^\beta_j \ddl{x^j_\beta}{x^i_\alpha}\,.
\end{equation}
Further define the following Poisson brackets on the extended
phase space:
\begin{equation}
\label{eq:ghostPB}
\pb{\cc^i}{\cP_j}=\delta^i_j\,, \quad \pb{\cc^i}{\cc^j}=0 \,, \quad
\pb{\cP_i}{\cP_j}=0 \,,
\end{equation}
the brackets between ghosts and other variables
vanish and the brackets among $x,y,p$ keep
their form \eqref{eq:PB}.  If the momenta $p_i$ were still
transformed according to~\eqref{eq:transition}
the Poisson bracket relations would not be
invariantly defined.  In order to make them
invariant we modify the transformation properties
of the momenta $p_i$: in the overlapping
$U_\alpha \cap U_\beta$ of coordinate neighborhoods
the transition law~\eqref{eq:transition} is modified
by ghost contribution as follows
\begin{equation}
        \label{eq:p-transform}
p^\alpha_i=p^\beta_j \frac{\d x^j_\beta}{\d x^i_\alpha}
        +\phi^{\alpha \beta}_i
+\cP^\beta_l \cc_\beta^k
\frac{\d x^j_\alpha}{\d x^k_\beta}
        \frac{\d^2{x^l_\beta}}{\d x_\alpha^i \d x_\alpha^j}\,.
\end{equation}
One can easily check that the Poisson brackets~\eqref{eq:ghostPB}
preserve their form under the change of coordinates on $\manM$
and the corresponding change of other variables.  Thus the
extended Poisson bracket is globally defined.

\noindent
Let us explain the geometry of the extended phase space
constructed above. Let $\bar \rho$ be the pullback of the
symplectic potential $\rho$ from the base $\manM$ to the odd
tangent bundle $\Pi T \manM$ over $\manM$
(we view ghost variables $\cc^i$ as natural coordinates
on the fibres of $\Pi T \manM$ over $\manM$). It was shown
in section~\bref{subsec:FM} that
given a (locally defined) 1-form $\rho_\alpha$
in each coordinate neighborhood $U_\alpha$ of
manifold $\manM$ one can construct a modified cotangent
bundle $\mod\manM$ over $\manM$.  If in addition 1-form is such
that $d\rho_\alpha=d\rho_\beta$ in the intersection
$U_\alpha \cap U_\beta$ the modified cotangent bundle
is equipped with the canonical symplectic
structure.  Applying this construction
to the $\Pi T \manM$ with the (locally defined) 1-form $\bar\rho$
one arrives at the affine bundle $\modbar(\Pi T \manM)$.

\noindent
Now it is easy to see that the Poisson bracket~\eqref{eq:ghostPB}
is nothing but the canonical Poisson bracket on the modified
cotangent bundle $\modbar(\Pi T \manM)$.  While the variable $p_i$,
being canonically conjugated to the variable $x^i$,
has the transition law~\eqref{eq:p-transform}.

\noindent
Finally, the whole extended phase space $\E$ of the BFV formulation
of the converted system is the vector bundle
\begin{equation}
\label{eq:E}
\E=\modbar(\Pi T \manM)\oplus T\manM,
\end{equation}
with $y^i$ being the natural coordinates on the fibres of $T
\manM$ (here $\modbar(\Pi T \manM)$ is considered as the vector bundle over
$\manM$ and $\oplus$ denotes the direct sum of vector bundles.) It
goes without saying that ``functions'' on $\E$ are formal power
series in $y^i$.  In what follows we denote the
algebra of ``functions'' on the extended phase space
by $\formal{\E}$.

\noindent
According to the BFV quantization procedure
we prescribe ghost degrees to each the variable
\begin{equation}
\gh{\cc^i}=1\,, \qquad \gh{\cP_i}=-1\,, \qquad
\gh{x^i}=\gh{y^i}=\gh{p_i}=0\,.
\end{equation}
Thus the Poisson bracket carries vanishing ghost
number.  A ghost charge $\G$ can be realized as
\begin{equation}
  \label{eq:ghost-charge}
\G=\cc^i\cP_i  \,.
\end{equation}
Indeed
\begin{equation}
\{\G,\cc^i\}=\cc^i\,,\quad \{\G,\cP_i\}=-\cP_i\,, \quad
\{\G,x^i\}=\{\G,y^i\}=\{\G,p_i\}=0\,.
\end{equation}
The BRST charge of the converted system is given by
\begin{equation}
  \label{eq:Omega}
  \Omega=\cc^i \cT_i\,.
\end{equation}
It satisfies the nilpotency condition
\begin{equation}
\label{eq:BFV-master}
\{\Omega,\Omega\}=0\,.
\end{equation}
w.r.t. the extended Poisson bracket.  The BRST charge
$\Omega$ carries unit ghost number:
\begin{equation}
\label{eq:Omega-GN}
\pb{\G}{\Omega}=\Omega\,.
\end{equation}
Relations \eqref{eq:BFV-master} and \eqref{eq:Omega-GN}
are known as the BRST algebra.
A BRST observable is a function $A$ satisfying
\begin{equation}
\pb{\Omega}{A}=0\,, \qquad \gh{A}=0\,.
\end{equation}
A BRST observable of the form $\pb{\Omega}{B}$
is called trivial.  The algebra of the inequivalent observables
(i.e. quotient of all observables modulo trivial ones)
is thus the zero-ghost-number cohomology of the classical
BRST operator $\pb{\Omega}{\cdot}$.  The
Poisson bracket on the extended phase space
obviously determines the Poisson bracket in
the BRST cohomology.  Thus the space of inequivalent
BRST observables is a Poisson algebra.
\begin{fact}
The Poisson algebra of inequivalent observables of the
BFV theory with the BRST charge~\eqref{eq:Omega} and ghost
charge~\eqref{eq:ghost-charge} is isomorphic with the Poisson
algebra of functions on the symplectic manifold $\manM$.
\end{fact}
\begin{proof}
Let $\aA_0 \subset \formal{\E}$ be the algebra of functions
depending on $x,y,\cc$ only.  Any function from $\aA_0$ is
a pullback of some function on $\Pi T\manM \oplus T\manM$ to
the entire extended phase space
$\E$ \eqref{eq:E}.  Let also
$A=A(x,y,p,\cc,\cP)$ be a BRST observable. Then one can check
that there exist functions $a\in \aA_0$ and $\Psi \in \formal{\E}$ such that
\begin{equation}
A=a+\pb{\Omega}{\Psi}\,, \qquad  \gh{a}=0\,,\quad \pb{a}{\Omega}=0\,.
\end{equation}
As a matter of fact $a$ can not depend on $\cc^i$
as it has zero ghost number.  Finally, it follows from
Proposition~\bref{fact:extention} that
the Poisson algebra of function on $\manM$
is isomorphic with the Poisson algebra of zero-ghost-number
BRST invariant functions from $\aA_0$.
\end{proof}
Thus at the classical level the initial Hamiltonian
dynamics on $\manM$ is equivalently represented as the
BFV theory.

\section{Quantization and quantum observables}
In this section we find a Poisson subalgebra $\aA$ in the algebra
of functions on the extended phase space which contains
all the physical observables and the generators of the
BRST algebra.  Thus instead of quantizing the entire
extended phase space it is sufficient to quantize
just Poisson subalgebra $\aA$.  This subalgebra can
be easily quantized that results in a quantum BRST
formulation of the effective first class constrained
theory.  The quantum BRST observables of the constructed
system are isomorphic to the space of functions on
the symplectic manifold $\manM$.  This isomorphism
carries the star-multiplication from the algebra of
quantum observables to the algebra of functions
on $\manM$, giving thus a deformation quantization
of $\manM$.  It turns out that the star multiplication
of the quantum BRST observables is the fibrewise
multiplication of the Fedosov flat sections of the
Weyl algebra bundle over $\manM$. Finally we interpret
all the basic objects of the Fedosov deformation quantization
as those of the BRST theory.

\subsection{Quantization of the extended phase space}

\noindent
Consider a Poisson subalgebra $\aA \subset \formal{\E}$
which is generated by subalgebra $\aA_0$
(subalgebra of functions depending on $x,y,\cc$ only)
and the elements
\begin{equation}
\P=\cc^i\theta_i \equiv \cc^i(\rho_i(x)-p_i)\,,\quad
\G=\cc^i \cP_i\,, \qquad \P,\,\G\in\formal{\E}\,.
\end{equation}
The reason for considering $\aA$ is that $\aA$ is a minimal
Poisson subalgebra of $\formal{\E}$ which contains,
at least classically, all the BRST observables and
both BRST and ghost charges (recall that $\G$
is precisely a ghost charge while BRST charge $\Omega$
can be represented in the form $\Omega=\P+{\bar\Omega}$,
with $\bar\Omega$ being some element of $\aA_0$).  A
general homogeneous element $a$ of $\aA$ has the form
\begin{equation}
a=\P^m
\G^n a(x,y,\cc)\,, \quad m=0,1\,,\quad n=0,1,\ldots N\,,\quad
N=\dim{\manM}\,, \quad a\in \aA_0\,,
\end{equation}
Note that algebra $\aA$ is not free, it can be considered
as the quotient of the free algebra generated (as a supercommutative
algebra) by $\aA_0$ and the elements $\P,\G$ modulo the relations
\begin{equation}
\begin{array}{c}
\P^m \cc^{i_1}\,,\ldots\,,\cc^{i_{N-k-m+1}} \G^k=0\,, \\
N=\dim{\manM}\,,\qquad k=0,1,\ldots,N+1\,,\quad m=0,1.\,,\quad N-k-m+1
\geq 0\,.
\end{array}
\end{equation}
The definition of $\aA$ is in fact invariant in
the sense that it is independent of the choice of
the coordinates on $\manM$.  The basic Poisson
bracket relations in $\aA$ read as
\begin{equation}
\begin{array}{rclrcl}
\displaystyle
\pb{\P}{\P}&=&R+\omega\,, \quad &
\pb{\G}{\P}&=&\P\,, \\[3pt]
\pb{\P}{a}&=&\nabla a\,, \quad &
\pb{\G}{\G}&=&0 \,,\\[3pt]
\pb{\G}{a}&=&\cc^i\ddl{a}{\cc^i}\,, \quad &
\pb{a}{b}&=&\omega^{ij}\ddl{a}{y^i}\ddl{b}{y^j}\,,\\
\end{array}
\end{equation}
where $a(x,y,\cc)$ and $b(x,y,\cc)$ are arbitrary elements of $\aA_0$,
$\nabla=\cc^i\nabla_i$ is a covariant differential in $\aA_0$
and
\begin{equation}
R=\frac{1}{2} R_{kl;ij} \cc^i \cc^j \,y^k y^l\,, \quad
\omega=\omega_{ij} \cc^i \cc^j \,, \qquad R,\omega \in \aA_0\,,
\end{equation}
is the curvature of the covariant differential $\nabla$
and the symplectic form respectively.  It is
easy to see that $\aA$ is closed w.r.t. the Poisson
bracket and thus it is a Poisson algebra.

\noindent
There is almost obvious star product which realizes
deformation quantization of $\aA$ as a Poisson
algebra.  The explicit construction of the star product
in $\aA$ is presented in Appendix~\bref{Appendix}.  In
order to proceed with the BRST quantization of our system
we will actually engage the star multiplication
in $\aA_0 \subset \aA$ given by the Weyl star-product
\begin{equation}
\label{eq:Weyl-star-product} (a \star b)(x,y,\cc)= exp ( -\frac{i\hbar}{2}
\omega^{ij} \dl{y_1^i} \dl{y_2^j}) a(x,y_1,\cc)b(x,y_2,\cc)|_{y_1=y_2=y} \,,
\end{equation}
and the following commutation relations in $\aA$:
\begin{equation}
\begin{array}{rclrcl}
\label{eq:basic-relations}
\qcommut{\P}{a}&=&-i\hbar \nabla a\,, \quad &
\qcommut{\P}{f(\G)}&=&i\hbar \P \dl{\G} F\,, \\[5pt]
\qcommut{\P}{\P}&=&-i\hbar(R+\Omega) \,, \quad &
\qcommut{\G}{a}&=&-i\hbar  \cc^i \dl{\cc^i}a\,,
\end{array}
\end{equation}
for any element $a \in \aA_0$
and a function $f(\G)$ depending on $\G$ only.  In
what follows $\aA$ and $\aA_0$ considered
as the associative algebras with respect to
star-multiplication will be denoted by $\aA^q$
and $\aA_0^q$ respectively.


\subsection{The quantum BRST charge}
\noindent
At the classical level, all the physical observables and
generators of the BRST algebra (ghost charge ${\G}$ and the
BRST charge $\Omega$) belong to $\aA$.  Thus to perform the
BRST quantization of the first class constrained system
one may restrict himself by the quantum counterpart
$\aA^q$ of the Poisson algebra $\aA$.

\noindent
Consider relations of the BRST algebra
\footnote{Here we have introduced a separate notation
$\hat\G$ for the quantum ghost charge because it can
differ from the classical ghost charge $\G=\cc^i \cP_i$ by
an imaginary constant $\frac{i\hbar N}{2}$. This constant,
of course, can not contribute to the commutation relations.}
\begin{equation} \label{eq:main}
\qcommut{{\hat\Omega}}{{\hat\Omega}} \equiv
2{\hat\Omega}*{\hat\Omega}=0\,, \qquad
\qcommut{\hat\G}{\hat\Omega}={\hat\Omega}\,,
\qquad {\hat\Omega},{\hat\G} \in \aA^q \,.
\end{equation}
The first equation implies the nilpotency of
the adjoint action $D$ of $\hat\Omega$ defined by
\begin{equation}
D A=\frac{i}{\hbar}\qcommut{\hat\Omega}{A}\,,
\qquad a\in \aA^q\,,
\end{equation}
Note that $D$ preserves subalgebra
$\aA^q_0 \subset \aA^q$ and therefore $D$ can
be considered as an odd nilpotent differential
in $\aA^q_0$.

\noindent
Show the existence of the quantum BRST charge
satisfying~\eqref{eq:main} whose classical limit coincides with
classical BRST charge $\Omega$ from previous section. Instead of
finding $\hbar$-corrections to the classical BRST charge it is
convenient to construct $\hat\Omega$ at the quantum level from
the very beginning.  In order to formulate
boundary conditions to be imposed on $\hat\Omega$
and for the technical convenience we introduce a useful
degree~\cite{[FL],[Fedosov-JDG]}.  Namely, we prescribe
the following degrees to the variables
\begin{equation}
\begin{split}
  \label{eq:degrees}
  \deg{x^i}=\deg{\cc^j}=0\,,&\qquad \deg{p_i}=\deg{\cP_i}=2 \,,\\
  \deg{y^i}=1\,,& \qquad  \deg{\hbar}=2\,.
\end{split}
\end{equation}
The star-commutator in $\aA^q$
apparently preserves the degree.

\noindent
Let us expand $\hat\Omega$ into the sum of homogeneous components
\begin{equation}
  \label{eq:Omega-expansion}
  {\hat\Omega}=\sum^\infty_{r=0} \Omega^r\,,\qquad \deg{\Omega^r}=r\,.
\end{equation}

\noindent
Given a classical BRST charge $\Omega$ \eqref{eq:Omega} which starts
as $\Omega=\cc^i\rho_i-\cc^i p_i+\cc^i\tau^1_{ij}y^j
+\cc^i\tau^2_{ijl}y^jy^l+\ldots$
one can formulate boundary condition on the solution
of~\eqref{eq:main} as follows
\begin{equation}
\label{eq:boundary}
  \Omega^0=\cc^i \rho_i\,,\qquad
  \Omega^1=\cc^i \tau^1_{ij}y^j\,.\qquad
  \Omega^2=-\cc^i p_i+\cc^i \tau^2_{ijl}y^jy^l \,.
\end{equation}
\begin{fact}\label{fact:main}
Equations \eqref{eq:main} considered as those for $\hat\Omega$
has a solution satisfying boundary condition~\eqref{eq:boundary}.
\end{fact}
\begin{proof}
Eq. \eqref{eq:main} evidently holds in the lowest order
w.r.t.  degree~\eqref{eq:degrees} provided respective
classical BRST charge $\Omega$ satisfies
$\pb{\Omega}{\Omega}=0$.  At the classical level the
higher order terms in expansion of $\Omega$ w.r.t. $y$ do
not depend on the momenta $p_i,\cP_i$.  Thus these terms belong
to $\aA_0$.  It is useful to assume that the
same occurs at the quantum level:
\begin{equation}
\Omega^r\in \aA^q_0 \qquad r \geq 3\,.
\label{eq:A0assumption}
\end{equation}
In the $r+2$-th $(r \geq 2)$ degree
Eq. \eqref{eq:main} implies
\begin{equation}
\label{eq:qdetermin}
\delta\Omega^{r+1}+B^r=0\,,
\end{equation}
where the quantity $B^r$ is
defined by $\Omega^t,\,t \leq r$:
\begin{equation}
B^r=\frac{i}{2\hbar}\sum_{t=0}^{r-2}
\qcommut{\Omega^{t+2}}{\Omega^{r-t}}\,,
\qquad
\deg{B^r}=r\,,
\end{equation}
and $\delta\,:\,\aA^q_0\to \aA^q_0$ stands for
\begin{equation}
  \delta a = \frac{i}{\hbar}\qcommut{\Omega^1}{a}
  = \cc^i\tau^1_{ij}\omega^{jl}\dl{y^l} a\,,
  \qquad a \in \aA^q_0\,.
\end{equation}
Note that $\delta$ is obviously nilpotent in
$\aA^q_0$.  It follows from the nilpotency of $\delta$ that the
compatibility condition for the Eq.~\eqref{eq:qdetermin}
is $\delta B^{r}=0$.  In fact it is a sufficient
condition for Eq.~\eqref{eq:qdetermin} to admit
a solution.  Indeed, the cohomology of the differential
$\delta$ is trivial when evaluated on functions at least linear
in $\cc$.  To show it, we construct the ``contracting
homotopy '' $\delta^{-1}$.  Namely, let $\delta^{-1}$
be defined by its action on a homogeneous element
\begin{equation}
  a_{pq}=a_{i_1,\ldots,i_p\,;\,j_1,\ldots,j_q}(x)\,y^{i_1}\ldots y^{i_p}
\cc^{j_1}\ldots \cc^{j_q}\,,
\end{equation}
by
\begin{equation}
\begin{split}
  \delta^{-1}a_{pq}&=\frac{1}{p+q}y^i (K^{-1})^j_i
\dl{\cc^j}a_{pq}\,,  \qquad p+q \neq 0 \\
 \qquad \delta^{-1} a_{00}&=0\,,~
\end{split}
\end{equation}
where $(K^{-1})^j_i$ is inverse to
$K^i_j=\tau^1_{il}\omega^{lj}$.  For any $a=a(x,y,\cc)$ we have
\begin{equation}
a|_{y=\cc=0}+\delta\delta^{-1} a+\delta^{-1}\delta
 a=a\,. \qquad
\end{equation}
Since $B^r$ is quadratic in
$\cc$ then the 3-rd term vanishes and $\delta B^r=0$ implies
$B^r=\delta\delta^{-1}B^r$ which in turn implies that
Eq.~\eqref{eq:qdetermin} admits a solution.

\noindent
Let us show that the necessary
condition $\delta B^r=0$ is fulfilled.  To this end assume
$\Omega^r$ to satisfy~\eqref{eq:qdetermin}
for $r \leq s$. Thus the Jacobi identity
\begin{equation}
  \qcommut{\sum_{t=0}^{s}\Omega^t}
{\qcommut{\sum_{t=0}^{s}\Omega^t}{\sum_{t=0}^{s}\Omega^t}}=0\,
\end{equation}
implies in the $s+3$-th degree that $\delta B^{s}=0$.

\noindent
The particular solution
to \eqref{eq:qdetermin} for $r=s$ evidently reads as
\begin{equation}
\label{eq:solution}
  \Omega^{s+1}=-\delta^{-1}B^s\,
\end{equation}
Iteratively applying this procedure one can construct a
solution to Eq.~\eqref{eq:main} at least locally.  To
show that Eq.~\eqref{eq:main} admit a global solution we note
that operators $\delta$ as well as $\delta^{-1}$ are defined in
a coordinate independent way.  It implies that particular
solution~\eqref{eq:solution} does not depend on the choice
of the local coordinate system and thus it is a global solution.

\noindent
The quantum BRST charge constructed above obviously
satisfies
\begin{equation}
\label{eq:add-condition}
 \delta^{-1}\Omega^r=0\,,\qquad r \geq 3 \,,
\end{equation}
which can be considered as an additional condition on
the solution to the Eq.~\eqref{eq:main}.  One can
actually show that solution to the Eq.~\eqref{eq:main} is unique
provided condition~\eqref{eq:add-condition} is imposed
on $\hat\Omega$.

\noindent
Thus we have shown how to
construct quantum BRST charge associated to the first
class constraints~$\cT_i$.
\end{proof}

\noindent
The operator $\delta$ which is extensively used in the proof
plays crucial role in the BRST formalism.  In the case of
the first class constrained system, the counterpart of $\delta$
is known as the Koszule-Tate differential associated to the
constraint surface~\cite{[H-omega],[HTS]} while in the Lagrangian
BV quantization~\cite{[BV]} the respective counterpart
of $\delta$ is the Koszule-Tate differential associated
to the stationary surface.

\subsection{An algebra of the quantum BRST observables
and the star-multiplication}
Observables in the BFV quantization are
recognized as zero-ghost-number values
closed w.r.t. to adjoint action $D$ of
BRST charge $\hat\Omega$ modulo exact
ones; $\hat a$ is an observable iff
\begin{equation}
\label{eq:Observables-determin}
  D{\hat a} \equiv \frac{i}{\hbar}\qcommut{\hat\Omega}{\hat a}=0\,,
  \qquad
  \qcommut{\hat\G}{\hat a}=0\,.
\end{equation}
Two observables are said equivalent iff their
difference is $D$-exact.  The space of inequivalent
observables is thus the zero-ghost-number cohomology of $D$.

\noindent
Initially, the classical observables are the functions
on the symplectic manifold $\manM$.  Now $\manM$ is embedded
into the extended phase space $\E$ \eqref{eq:E}.  According to the
BFV prescription the quantum extension of the
initial observable $a$ is an operator (symbol in our case)
$\hat a$ of the quantum converted system that is the solution
to the Eqs.~\eqref{eq:Observables-determin}
subjected to the boundary condition
\begin{equation}
\label{eq:bc}
{\hat a}|_{y=0}=a_0(x)\,.
\end{equation}
Let us consider the algebra of quantum BRST observables in
$\aA^q$. First study observables in $\aA^q_0 \subset \aA^q$.
\begin{fact}
\label{fact:observ-existence}
Equations \eqref{eq:Observables-determin} have a unique solution
belonging to $\aA_0^q$ for each initial observable $a_0=a_0(x)$.
\end{fact}
\begin{proof}
Consider an expansion of $\hat a$ in the homogeneous components
\begin{equation}
    \label{eq:a-expuncion}
\hat a=\sum_{r=0}^\infty a^r\,,\qquad \deg{a^r}=r\,.
\end{equation}
The boundary condition~\eqref{eq:bc} implies
$a^0=a_0(x)$.  Eq.~\eqref{eq:Observables-determin}
obviously holds in the first degree.  In the higher
degrees we have
\begin{equation}
  \label{eq:a-determin}
\delta a^{r+1}+B^r=0\,,
\end{equation}
where $B^r$ is given by
\begin{equation}
B^r=\frac{i}{\hbar}\sum_{t=0}^{r-2}
\qcommut{\Omega^{t+2}}{a^{r-t}}\,,
\qquad
\deg{B^r}=r\,,
\end{equation}
and $\Omega^t$ are terms of the expansion~\eqref{eq:Omega-expansion}
of ${\hat\Omega}$ w.r.t. degree.  Similarly to the proof of the
Proposition~\bref{fact:main} the necessary and sufficient condition
for~\eqref{eq:a-determin} to admit solution is $\delta B^r=0$.  To
show that the condition holds indeed assume that
Eqs.~\eqref{eq:a-determin} hold for all $r\leq s$.  Then consider
the identity
\begin{equation}
  \qcommut{{\hat\Omega}}{\qcommut{{\hat\Omega}}{\hat a}}=0\,.
\end{equation}
Excluding contribution of degree $s+3$ we arrive at
\begin{equation}
\delta B^s=0\,.
\end{equation}
The particular solution to the Eq.~\eqref{eq:a-determin}
for $r=s$ is
\begin{equation}
  a^{s+1}=-\delta^{-1}B^s\,.
\end{equation}
Iteratively applying this procedure one arrives at the particular
solution to the Eq.~\eqref{eq:Observables-determin} satisfying
boundary condition ${\hat a}|_{y=0}=a_0(x)$.

\noindent
Finally let us show the uniqueness.  Taking into account that
$a^{s+1}$ belongs to $\aA^q_0$ and $\gh{a^{s+1}}=0$ we conclude
that $a^{s+1}$ does not depend on $\cc$.  Thus the general
solution to the equation~\eqref{eq:a-determin} is given by
\begin{equation}
  \label{eq:general}
a^{s+1}=-\delta^{-1}B^s+C^{s+1}(x,\hbar)\,.
\end{equation}
It is easy to see that the boundary condition requires
$C^{s+1}(x,\hbar)=0$ (recall that $(\delta^{-1}B^n)|_{y=0}=0$.)
\end{proof}

\noindent
Since equation~\eqref{eq:Observables-determin} is linear
it has a unique solution $\hat a$ satisfying the
boundary condition ${\hat a}_{y=0}=a_0(x,\hbar)$
even if the initial observable $a_0$ was allowed
to depend formally on $\hbar$.  It follows from
the Proposition~\bref{fact:observ-existence}
that the space of inequivalent quantum observables coincides
with the space of classical observables (functions on $\manM$)
tensored by formal power series in $\hbar$.  In other words, the
space of zero-ghost-number cohomology
of $D$ evaluated in $\aA^q_0$ is isomorphic with
$\func\manM\tensor [[\hbar]]$, where $\func\manM$ is the
algebra of functions on $\manM$ and $[[\hbar]]$ denotes
the space of formal power series
in $\hbar$.  In fact even a stronger statement holds
\begin{thm}
\label{thm:Q-observables}
The space of inequivalent quantum BRST observables, i.e.
the zero-ghost-number cohomology evaluated in $\aA^q$,
is isomorphic to $\func{\manM} \tensor [[\hbar]]$.
\end{thm}
A proof follows from observation that each zero-ghost-number
cohomology class from $\aA^q$ has a representative
in $\aA^q_0$.  Further, Proposition~\bref{fact:observ-existence}
implies that the representative is unique and provides us
with the explicit isomorphism between $\func\manM \tensor [[\hbar]]$
and the space of inequivalent quantum observables.

\noindent
Since the BRST differential
$D=\frac{i}{\hbar}\qcommut{\hat\Omega}{\cdot\,}$
is an inner derivation in $\aA^q$ then the
star multiplication in $\aA^q$ determines
the star multiplication in the space of
quantum BRST cohomology.  Making use of isomorphism
from~\bref{thm:Q-observables} one can equip
$\func\manM\tensor [[\hbar]]$ with the associative
multiplication, determining thereby a star product on
$\manM$.  As the algebra of quantum BRST observables is
a deformation of the Poisson algebra of
classical ones, which in turn is isomorphic with the
Poisson algebra $\func{\manM}$, the star-product on
$\manM$ satisfies standard correspondence conditions
\begin{equation}
{\hat a}\star {\hat b}|_{y=\hbar=0}=a_0 b_0\,, \qquad
  \frac{i}{\hbar}\qcommut{\hat a}{\hat b}|_{y=\hbar=0}
  =\pb{a_0}{b_0}_{\manM}\,.
\end{equation}
Here $\hat a$ and $\hat b$ are the symbols obtained by means
of Proposition~\bref{fact:observ-existence} starting from
$a_0,b_0\in\func\manM$.

\subsection{BFV-Fedosov correspondence}
To establish the correspondence with the Fedosov construction
of the star product we note that the quantum algebra
$\aA_0^q$ consisting of functions\footnote{Recall that in
the BRST approach functions of auxiliary variables
$y$ are formal power series in $y$} of $x,y,\cc$ is precisely
the algebra of sections of the Weyl algebra bundle from
\cite{[Fedosov-JDG]} provided one identifies ghosts $\cc^i$ with the
basis 1-forms $dx^i$.

\noindent
Let us consider the quantum BRST charge $\hat\Omega$
corresponding to the boundary condition~\eqref{eq:tau-particular}.  An
adjoint action of $\hat\Omega$ on $\aA^q_0$
\begin{equation}
Da \equiv \frac{i}{\hbar}\qcommut{\hat\Omega}{a}=
(\cc_i\nabla_i-\cc^i\dl{y^i})a+\frac{i}{\hbar}
\qcommut{\sum^\infty_{t=3} \Omega^t}{a}\,,
\quad a\in\aA^q_0\,,
\end{equation}
is precisely the Fedosov connection in the Weyl algebra
bundle.  Indeed, in the Fedosov-like notations
\begin{equation}
  \label{eq:fedosov}
  \delta=\cc^i\dl{y^i}\,,\qquad
  \partial=\cc^i \nabla_i\,,\qquad
  r=\sum_{t=3}^\infty \Omega^t \in \aA_0\,,
\end{equation}
the star-commutator with quantum BRST charge $\hat\Omega$
can be rewritten as
\begin{equation}
Da \equiv \frac{i}{\hbar}\qcommut{\Omega}{a}=
(\partial-\delta)a+\frac{i}{\hbar}\qcommut{r}{a}\,,
\end{equation}
that makes transparent identification of the
$\frac{i}{\hbar}\qcommut{\hat\Omega}{\cdot\,}$ and the
Fedosov connection in the Weyl algebra bundle.  In
particular, the zero-curvature condition is precisely
the BFV quantum master equation
$\qcommut{\hat\Omega}{\hat\Omega}=0$.

\noindent
It follows from Theorem~\bref{thm:Q-observables} that
each equivalence class of the quantum BRST observables has
a unique representative in $\aA^q_0$.  Thus the inequivalent
quantum BRST observables are the flat sections of the
Weyl algebra bundle.  In its turn the star
multiplication~\eqref{eq:Weyl-star-product} of quantum
BRST observables (considered as the BRST invariant
functions from $\aA_0^q$) is nothing but the Weyl
product of the Fedosov flat sections of Weyl algebra bundle.

\noindent
There is a certain distinction between BRST and Fedosov
quantization.  Unlike the Fedosov Abelian connection,
the adjoint action of the BRST charge can be realized
as an inner derivation of the associative algebra
$\aA^q$.  In particular, in the BRST approach the
covariant differential $D$ is strictly flat.

\section{Conclusion}

\noindent
Summarize the results of this paper.  First we construct a global
embedding of a general symplectic manifold $\manM$
into the modified cotangent bundle $\mod\manM$ as a second class
constrained surface.  Then we have elaborated globally defined procedure
which converts the second class constrained system into the first
class one that allows to construct the BRST description
for the Hamiltonian dynamics in the original
symplectic manifold.  We have explicitly established
the structure of the classical BRST
cohomology in this theory and perform a straightforward quantum
deformation of the classical Poisson algebra which contains all the
observables and the BRST algebra generators.  As all the values
on the original symplectic manifold are identified with the
observables of the BRST theory, we have thus quantized the
general symplectic manifold.  Finally, we establish a detailed
relationship between the quantum BFV-BRST theory of the symplectic
manifolds and Fedosov's deformation quantization.

\noindent
The construction of the BRST embedding of the second
class constrained theory, being done by means of the
cohomological technique, allows to recognize the
conversion procedure as some sort of deformation of the classical
Poisson algebra of the second class system. This deformation has an essential
distinction from that one which is usually studied in relation to switching
on the interactions in classical gauge theories \cite{henneaux-deform},
although the cohomological technique is quite similar. As
soon as {\it classical} deformation has been performed, the problem of the
{\it quantum} deformation becomes transparent in the theory.  Thus in the
BRST approach a part of the deformation quantization problem is transformed,
in a sense, in the problem of another deformation, classical in essence,
while the quantization itself is almost obvious in the classically deformed
system.

\subsection*{Acknowledgments}
We are grateful to I.A.~Batalin for fruitful discussions
on various problems considered in this paper. We also wish
to thank V.A.~Dolgushev, A.V.~Karabegov, A.M.~Semikhatov,
A.A.~Sharapov, I.Yu.~Tipunin and I.V.~Tyutin.  The work of
MAG is partially supported by the RFBR grant 99-01-00980,
INTAS-YSF-98-156, Russian Federation President Grant~99-15-96037
and Landau Scholarship Foundation, Forschungszentrum
J\"ulich.  The work of SLL is supported by the
RFBR grant 00-02-17956.

\appendix
\section{Star-multiplication in $\aA$}\label{Appendix}
Here we present an explicit form of the star product
in the Poisson algebra $\aA$. Recall, that
$\aA$ is a Poisson subalgebra of $\formal{\E}$
generated by $\aA_0$ (Poisson algebra of functions depending of
$x,y,\cc$ only) and the elements
$\P=\cc^i\theta_i,\,\G=\cc^i \cP_i$. The star product in $\aA_0$
could be defined by the Weyl
multiplication~\eqref{eq:Weyl-star-product}.  As for general
elements of $\aA$, let us consider first $\P$-independent
ones.  For the general  $\P$-independent elements $A(x,y,\cc,\G)$
and $B(x,y,\cc,\G)$ we postulate
\begin{equation}
\label{eq:P-independent}
\begin{array}{c}
A(x,y,\cc,\G)\star B(x,y,\cc,\G)=\bigl[
exp (-i\hbar(\G\dl{\G_1}\dl{\G_2}+
\cc^i\dl{\cc^i_2}\dl{\G_1}))\\
A(x,y,\cc_1,\G_1)\star B(x,y,\cc_2,\G_2)
\bigr] {\bigg|}_{\cc_1=\cc_2=\cc,\, \G_1=\G_2=\G,}
\end{array}
\end{equation}
where the $\star$ in the r.h.s. is the Weyl
multiplication~\eqref{eq:Weyl-star-product}
acting on $y$ only.  Finally, taking into account commutation
relations~\eqref{eq:basic-relations} for the
$\P$-dependent elements one can choose
\begin{equation}
\label{eq:P-dependent}
\begin{array}{rcl}
(\P A)\star(B)&=&\P (A\star B)\,,\\[5pt]
(A)\star(\P B)&=&(-1)^{\p{A}}\bigl[
\P (A\star B)
-i \hbar \P ((\dl{\G}A)\star B)\\ [5pt]
&&
+i\hbar ((\nabla A)\star B)
-(i\hbar)^2 (\nabla \dl{\G} A)\star B \bigr] \,,\\ [5pt]
(\P A)\star(\P B)&=&(-1)^{\p{A}}\bigl[
\frac{i\hbar}{2} (R+\omega) \star A \star B
+\frac{(i \hbar)^2}{2} (R+\omega) \star (\dl{\G} A)\star B\\[5pt]
&&
+i\hbar \P ((\nabla A) \star B)
-(i\hbar)^2 \P ((\nabla \dl{\G} A)\star B) \bigr]\,,\\
\end{array}
\end{equation}
where $A$ and $B$ are general $\P$-independent elements
from $\aA$ and the star-product in the right hand sides of
Eqs.~\eqref{eq:P-dependent} is that given
by~\eqref{eq:P-independent}.  The associativity
of multiplication~\eqref{eq:P-independent} and \eqref{eq:P-dependent}
can be verified directly.

\noindent
This star multiplication can be thought of
as that of $\P,x,y,\cc,\G$-symbols
which is also a Weyl symbol w.r.t. $y$-variables.  If
one considered the star-product in $\aA$ as
that on functions explicitly depending on $p,\cP$
then it would correspond to the $p,x,y,\cc,\cP$-symbol.

\end{document}